\documentclass[11pt]{article}

\usepackage[margin=1in]{geometry}
\usepackage{amsmath,amssymb,bm,mathtools}
\usepackage{graphicx,booktabs}
\usepackage{float}
\usepackage{microtype}
\usepackage[hidelinks]{hyperref}
\usepackage{url}
\usepackage{enumitem}
\graphicspath{{./}}
\hypersetup{pdftitle={Prediction Accuracy Can Select the Scaling Law of Required Precision},pdfauthor={Song-Ju Kim}}

\newcommand{\cO}{\mathcal{O}}
\newcommand{\Th}{\Theta}
\newcommand{\eps}{\varepsilon}
\newcommand{\Torus}{\mathbb{T}}

\title{Prediction Accuracy Can Select the Scaling Law of Required Precision}
\author{Song-Ju Kim\\
\small SOBIN Institute LLC, Kawanishi, Hyogo, Japan\\
\small \texttt{kim@sobin.org}}
\date{September 30, 2026}

\begin{document}
\maketitle

\begin{abstract}
How much precision must be allocated, and to which uncertainty channels, to meet a prediction requirement?
We formulate a minimum one-shot specification cost over initial state variables and persistent forcing parameters.
The central result is that prediction accuracy can select the horizon-scaling law of this resource by activating different physical uncertainty channels.
Within any common uniformly scale-separable regime with the same active channels, changing a fixed tolerance changes the required precision by only $\cO(1)$ bits.
An exactly solvable multiscale toral system shows that one fixed dynamics and target can nevertheless switch from
$B_N(\eps_c)=\Th(\log N)$ to $B_N(\eps_f)=\Th(N)$ when finer accuracy activates an expanding channel.
A forced Lorenz--84 calculation then realizes the same channel-switching mechanism numerically over finite horizons in a single two-scale diagnostic field: a sampled post-transient eddy envelope supports a coarse branch controlled by the phase rate of a large-scale forcing, whereas finer accuracy activates exponentially sensitive atmospheric degrees of freedom and produces an approximately linear increase of the linearized one-shot cost over the resolved window.
Thus accuracy can determine not only how much predictive precision is required, but which physical channel controls its growth with horizon.
\end{abstract}

Prediction error can be characterized by Lyapunov growth, finite-size predictability, or the largest admissible initial error for a prescribed lead time and tolerance \cite{Boffetta1998,Cao2018}.  Numerical-precision studies also connect retained bits to forecast lead time \cite{Cooper2020}.  Dynamical-systems theory distinguishes exponential from polynomial and more general orders of orbit separation \cite{BrinKatok1983,Marco2013,CorreaPujals2023}, while estimation and restoration entropies relate trajectory reconstruction to sustained communication rates \cite{MatveevPogromsky2019,SibaiMitra2023}.  Adaptive-observation methods identify sensitive directions or regions for a specified forecast objective \cite{Palmer1998}.  These approaches quantify sensitivity, admissible error, orbit complexity, or continuing information requirements from complementary viewpoints.  Here the object is an inverse allocation problem: for a declared prediction product, horizon, and accuracy, what is the minimum \emph{one-shot resolution allocation} across specified physical state and forcing channels, which channels receive that precision, and how does the minimum scale with horizon?

A related finite-resource formulation optimizes the anisotropic geometry of local state uncertainty when its volume is fixed \cite{Kim2026RIPG}.  Here the resource itself is not fixed.  The prediction requirement determines the minimum necessary precision.

Let $T$ be the prediction horizon, $\eps$ the allowed error, and $\bm q_0=(q_{01},\ldots,q_{0d})$ a nominal $d$-dimensional predictive specification.  Its coordinates may include initial state variables and persistent external parameters known only to finite precision.  A rectangular uncertainty cell is
\[
\mathcal U(\bm q_0,\bm\rho)=
\{\bm q:|q_i-q_{0i}|\le \rho_i\},
\]
where $\rho_i$ is the uncertainty half-width in coordinate $q_i$.  Let $H(\bm q;t)$ denote the predicted observable or field at time $t$, and let $\|\cdot\|$ be the norm defining the prediction requirement.  The horizon-wide minimax radius is
\begin{equation}
R_T(\bm q_0,\bm\rho)=
\inf_{\widehat H}
\sup_{\bm q\in\mathcal U(\bm q_0,\bm\rho)}
\sup_{0\le t\le T}
\|H(\bm q;t)-\widehat H(t)\|,
\label{eq:radius}
\end{equation}
where the estimate $\widehat H$ may use the model, $\bm q_0$, and the uncertainty cell but not the unknown true $\bm q$.  For fixed reference half-widths $r_i$, with $0<\rho_i\le r_i$, define
\begin{equation}
B_T(\bm q_0,\eps)=
\inf_{R_T(\bm q_0,\bm\rho)\le\eps}
\sum_{i=1}^d\log_2\frac{r_i}{\rho_i}.
\label{eq:Bdef}
\end{equation}
This one-shot cost is attached to a declared sensing architecture rather than to a continuing communication channel.  The physical coordinates and reference widths $r_i$ specify that architecture.  A prediction task may also include a fixed admissible-set prior $\mathcal K$, representing physical information available before any additional refinement; then $\mathcal U(\bm q_0,\bm\rho)$ is replaced by $\mathcal U(\bm q_0,\bm\rho)\cap\mathcal K$.  Such a prior is part of the task definition and is held fixed as $T$ and $\eps$ vary.  Unless stated otherwise, $\mathcal K$ is the full declared state domain.  A fixed rescaling of a coordinate together with its uncertainty and reference widths leaves the corresponding logarithmic contribution unchanged.  Below we suppress the explicit $\bm q_0$ argument when no ambiguity arises.

There is a simple obstruction to an accuracy-induced class change.  Suppose that, throughout one common local regime,
\begin{align}
c_-\mathcal{G}(T)\Phi(\bm\rho)
&\le R_T(\bm q_0,\bm\rho)
\le c_+\mathcal{G}(T)\Phi(\bm\rho),\nonumber\\
\Phi(s\bm\rho)&=s^p\Phi(\bm\rho).
\label{eq:separable}
\end{align}
where $\mathcal{G}(T)$ contains the horizon dependence, $\Phi$ the uncertainty-cell dependence, $p>0$ is its homogeneity degree, and $0<c_-\le c_+$ are independent of $T$.  If the same $d$ active coordinates apply at two fixed accuracies $0<\eps_1<\eps_2$, radial rescaling of a near-optimal cell gives
\begin{equation}
0\le B_T(\eps_1)-B_T(\eps_2)
\le \frac{d}{p}\log_2\!\left(\frac{c_+\eps_2}{c_-\eps_1}\right).
\label{eq:notransition}
\end{equation}
Thus accuracy can select a different asymptotic class only when the common scale-separable description fails or the active predictive channels change.  A proof is given in Appendix~\ref{app:no-transition}.  This identifies the mechanism of interest: accuracy acts as a channel selector.  At the level of scaling, a channel that must be resolved as $T^{-\alpha}$ contributes $\alpha\log_2 T+\cO(1)$ bits, whereas a newly activated channel that must be resolved as $e^{-\lambda T}$ contributes $(\lambda/\ln2)T+\cO(1)$.  The exact model below realizes this logarithmic-to-linear switch analytically; the Lorenz--84 calculation tests the same channel-activation mechanism in a coupled continuous flow.

That possibility occurs exactly in the toral map
\begin{align}
u_{n+1}&=u_n+v_n\pmod 1,\qquad v_{n+1}=v_n\pmod 1,\nonumber\\
z_{n+1}&=mz_n+u_n\pmod 1,\qquad m\ge2.
\label{eq:torus}
\end{align}
where $n$ is the discrete prediction step and $m$ is the expansion factor of the $z$ channel.  For the fixed target $(u,z)$ with the metric specified in Appendix~\ref{app:torus}, a coarse tolerance can leave the bounded $z$ coordinate unresolved while still requiring $v$ to precision $\propto N^{-1}$, giving
\[
B_N(\eps_c)=\log_2 N+\cO(1).
\]
A finer tolerance activates the expanding $z$ channel; for all sufficiently large $N$, the exact optimum satisfies
\[
B_N(\eps_f)=3N\log_2m+\cO(1).
\]
Hence one fixed system and target have two unbounded resource laws, $\Th(\log N)$ and $\Th(N)$, selected only by accuracy.

We next test the mechanism in the Lorenz--84 model of atmospheric circulation \cite{Lorenz1984}, using a periodically forced variant motivated by earlier driven Lorenz--84 studies \cite{Broer2002}:
\begin{align}
\dot X&=-Y^2-Z^2-aX+a[F_0+F_1\cos\theta],\nonumber\\
\dot Y&=XY-bXZ-Y+G,\nonumber\\
\dot Z&=bXY+XZ-Z,
\label{eq:L84}
\end{align}
with $\theta(t)=\theta_0+ct$.  Here $X$ represents the large-scale westerly circulation, $Y$ and $Z$ are the cosine and sine amplitudes of a large-scale eddy mode, $a$ and $b$ are the standard damping/advection parameters, $F_0$ is the mean symmetric thermal forcing, $F_1$ its periodic modulation amplitude, and $G$ the asymmetric thermal forcing.  The phase $\theta$ locates the external large-scale pattern and $c$ is its persistent phase rate.  Including $\dot c=0$ places the persistent phase rate in the tangent calculation while leaving it physically an external forcing specification.

To compare the large-scale forcing and intrinsic eddy response without allowing cancellation between them, we use one scalar two-scale diagnostic field
\begin{equation}
H(\lambda,t)=A_P\cos[\lambda-\theta(t)]
+Y(t)\cos(2\lambda)+Z(t)\sin(2\lambda),
\label{eq:field}
\end{equation}
where $\lambda\in[0,2\pi)$ is a periodic spatial coordinate and $A_P$ is the large-scale pattern amplitude.  Error is measured by the normalized spatial norm
$\|h\|_2=[\pi^{-1}\int_0^{2\pi}|h(\lambda)|^2d\lambda]^{1/2}$.
The distinct Fourier modes provide an orthogonal diagnostic encoding of the two physical scales, so errors in the imposed large-scale pattern and intrinsic eddy response cannot cancel in the target norm.

We use
$a=0.25$, $b=4$, $F_0=8$, $F_1=0.5$, $G=1$, $c=0.25$, and $A_P=3$.
For the coarse construction, let $\overline R_e$ denote a uniform bound on $\sqrt{Y^2+Z^2}$ over the declared post-transient atmospheric set $\mathcal K_{\rm atm}$.  Whenever $\overline R_e<\eps_c<A_P$, the eddy channel can remain at its reference resolution while the $k=1$ forcing channel must still be refined.  If $L_T=\rho_\theta+T\rho_c$ is the phase half-width accumulated from initial phase and phase-rate uncertainties, orthogonality gives the sufficient condition
\begin{equation}
\overline R_e^{\,2}+A_P^2\sin^2L_T\le\eps_c^2.
\label{eq:coarsecondition}
\end{equation}
A long post-transient trajectory gives the empirical maximum $R_e^{\rm samp}=2.429$.  Using this sampled envelope in the conditional formula with $\eps_c=2.70$ gives $L_T\le L_c=0.404$ rad.  Maximizing $\rho_\theta\rho_c$ under $\rho_\theta+T\rho_c\le L_c$ gives $\rho_c\propto T^{-1}$ and hence a logarithmic forcing-channel cost.  Projection onto the orthogonal $k=1$ mode gives the matching necessity $\rho_c=\cO(T^{-1})$.  Thus any uniform post-transient eddy bound satisfying $\overline R_e<\eps_c$ produces a logarithmic coarse forcing branch; the curve shown below is its finite-data realization obtained from $R_e^{\rm samp}$.

For the fine tolerance $\eps_f=0.35$, eddy uncertainty is active.  Let $M(t)=D\bm q(t)/D\bm q(0)$ be the tangent map for $\bm q=(X,Y,Z,\theta,c)$ and define the horizon sensitivity of coordinate $i$ by
\begin{equation}
\Gamma_i(T)=\sup_{0\le t\le T}
\sqrt{[A_PM_{\theta i}(t)]^2+M_{Yi}(t)^2+M_{Zi}(t)^2}.
\label{eq:gamma}
\end{equation}

\begin{figure}[t]
\centering
\includegraphics[width=0.78\linewidth]{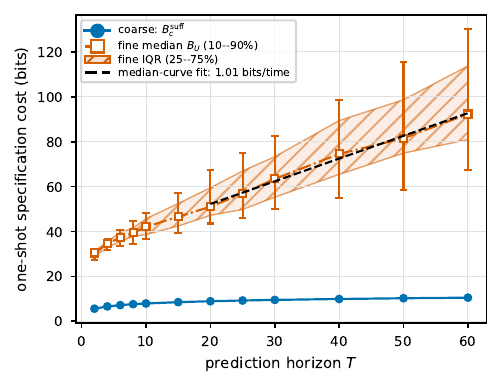}
\caption{Accuracy-selected precision scaling in the forced Lorenz--84 realization.  Filled circles and a solid line show the analytic sufficient coarse forcing-channel cost implied by Eq.~\eqref{eq:coarsecondition}.  Open squares with a dash-dotted line show the median fine linearized upper bound $B_U$ from Eq.~\eqref{eq:bounds} over 64 post-transient starts; the hatched band is the 25--75\% range and whiskers the 10--90\% range.  The black dashed segment is the finite-window fit to the median fine curve for $T\ge20$.  The $1.01$ bits/time value is obtained by fitting the ensemble-median curve; the median of the 64 individually fitted slopes is $0.96$ bits/time (IQR $0.70$--$1.47$).  The intervals display variation across starting states.  Marker, line style, and hatching keep all curves distinguishable in grayscale.  The lower bound $B_L$ differs from $B_U$ by at most $5\log_2 5=11.61$ bits, independent of $T$, so the two linearized bounds have the same asymptotic horizon class whenever that class is defined.}
\label{fig:scaling}
\end{figure}

The coarse and fine resource curves are summarized in Fig.~\ref{fig:scaling}; the fine linearized bound used in that comparison is derived next.

For the exact box-optimal cost $B^*_{\rm lin}$ of the local linearized problem, coordinatewise necessity and the triangle inequality give the computable bracket
\begin{align}
B_L&=\sum_i\left[\log_2\frac{r_i\Gamma_i}{\eps_f}\right]_+
\le B^*_{\rm lin},\nonumber\\
B^*_{\rm lin}&\le
\sum_i\left[\log_2\frac{d r_i\Gamma_i}{\eps_f}\right]_+=B_U.
\label{eq:bounds}
\end{align}
where $[x]_+=\max(x,0)$ and $B_U-B_L\le d\log_2d$, independent of $T$.  Thus either bound determines the same horizon-scaling class.

Figure~\ref{fig:scaling} uses $d=5$, $r_i=1$ in the declared model-coordinate units, and 64 post-transient starting states sampled along one long reference trajectory.  The median fine upper-bound curve has a fitted finite-window slope of $1.01$ bits per model-time unit over $T\ge20$; fitting each start separately gives a slope median $0.96$ and interquartile range $0.70$--$1.47$.  The slowly growing coarse curve and the approximately linear fine curve make the channel switch visible directly: finer accuracy changes not merely the amount of required precision but the physical channel that controls its horizon dependence.  The hatched interquartile band and 10--90\% whiskers in Fig.~\ref{fig:scaling} display the substantial dynamical variation across starting states; marker shapes and line styles distinguish all plotted quantities independently of color.  Direct nonlinear finite perturbations, parameter scans, the lower/upper bound comparison, and an autonomous closure are reported in Appendices~\ref{app:fine}--\ref{app:autonomous}.

The operational distinction from existing sensitivity and entropy constructions is concrete.  Targeted observations identify sensitive directions or locations for a chosen objective and lead time \cite{Palmer1998}; entropy and data-rate theories characterize orbit complexity or rates needed for sustained state reconstruction \cite{MatveevPogromsky2019,SibaiMitra2023}.  Equation~\eqref{eq:Bdef} instead returns a one-shot design quantity: the minimum initial resolution allocation across declared physical channels that meets a horizon-wide accuracy.  In the Lorenz--84 realization, changing accuracy shifts the precision-limiting channel from the phase rate of the imposed large-scale pattern to the exponentially sensitive atmospheric state.  In the exact model, the same activation mechanism changes the asymptotic law itself from $\Th(\log T)$ to $\Th(T)$.

Prediction accuracy is therefore not merely a numerical threshold applied after a resource law has been fixed.  It can decide which physical uncertainty must be resolved and, when those channels carry different growth laws, which horizon-scaling law governs the required precision.

\appendix

\section{Local minimax precision resource}\label{app:resource}
For a nominal predictive specification $\bm q_0=(q_{01},\ldots,q_{0d})$ and coordinate half-widths $\bm\rho=(\rho_1,\ldots,\rho_d)$, define
\[
\mathcal U(\bm q_0,\bm\rho)=\{\bm q:|q_i-q_{0i}|\le\rho_i\}.
\]
If the task includes a fixed admissible-set prior $\mathcal K$, the uncertainty set is $\mathcal U_{\mathcal K}=\mathcal U\cap\mathcal K$.  This separates pre-existing physical information, such as membership in a bounded post-transient set, from the additional local resolution being purchased.  For a prediction target $H(\bm q;t)$, prediction horizon $T$, and target norm $\|\cdot\|$, the minimax radius is
\[
\begin{aligned}
R_T(\bm q_0,\bm\rho)=\inf_{\widehat H}\;&\sup_{\bm q\in\mathcal U_*(\bm q_0,\bm\rho)}\\[-2pt]
&\sup_{0\le t\le T}\|H(\bm q;t)-\widehat H(t)\|,
\end{aligned}
\]
where $\mathcal U_* = \mathcal U$ without an admissible-set prior and $\mathcal U_* = \mathcal U_{\mathcal K}$ when such a prior is declared.
The estimator $\widehat H$ may depend on the model, the nominal specification, and the declared uncertainty cell, but not on the unknown true $\bm q$ inside that cell.  For fixed reference half-widths $r_i$ and $0<\rho_i\le r_i$, the additive one-shot precision cost is
\[
B(\bm\rho)=\sum_{i=1}^d\log_2(r_i/\rho_i),
\qquad
B_T(\bm q_0,\eps)=\inf_{R_T\le\eps}B(\bm\rho).
\]
The physical coordinate channels, their reference scales, and any admissible-set prior are part of the declared sensing architecture.  A fixed rescaling of one coordinate together with its uncertainty and reference widths leaves the corresponding logarithmic cost unchanged.  More general coordinate mixing changes how precision is allocated because it represents a different sensor architecture; horizon-scaling comparisons in this paper are always made with that architecture fixed.  Positive coordinate weights $w_i$ can replace the equal weights used here; fixed weights change coefficients but not the logarithmic-versus-linear distinction.

\section{No-transition lemma}\label{app:no-transition}
Assume that for a common local regime and all sufficiently large $T$,
\[
c_-\mathcal{G}(T)\Phi(\bm\rho)\le R_T(\bm q_0,\bm\rho)\le c_+\mathcal{G}(T)\Phi(\bm\rho),
\]
where $0<c_-\le c_+$ do not depend on $T$ and $\Phi(s\bm\rho)=s^p\Phi(\bm\rho)$ for $s>0$, $p>0$.  Let $0<\eps_1<\eps_2$ and assume the same $d$ coordinates remain in the common regime for both accuracies.

Monotonicity gives $B_T(\eps_1)\ge B_T(\eps_2)$.  Let $\bm\rho_2$ be arbitrarily close to optimal for $\eps_2$.  Since
$c_-\mathcal{G}\Phi(\bm\rho_2)\le R_T(\bm\rho_2)\le\eps_2$,
we have $\Phi(\bm\rho_2)\le\eps_2/(c_-\mathcal{G})$.  Define
\[
s=\left(\frac{c_-\eps_1}{c_+\eps_2}\right)^{1/p}<1.
\]
Then
\[
R_T(s\bm\rho_2)\le c_+\mathcal{G}s^p\Phi(\bm\rho_2)\le\eps_1.
\]
Shrinking all active half-widths by $s$ raises the additive cost by at most $d\log_2(1/s)$.  Taking the infimum yields
\[
0\le B_T(\eps_1)-B_T(\eps_2)
\le\frac{d}{p}\log_2\left(\frac{c_+\eps_2}{c_-\eps_1}\right).
\]
The right-hand side is independent of $T$.  Therefore two fixed accuracies cannot have different horizon-scaling classes while both remain in one uniformly scale-separable regime with the same active coordinates.

\section{Exact toral transition}\label{app:torus}
Consider $(u,v,z)\in\Torus^3$ with
\begin{align}
u_{n+1}&=u_n+v_n\pmod1,\\
v_{n+1}&=v_n\pmod1,\\
z_{n+1}&=mz_n+u_n\pmod1,\qquad m\ge2.
\end{align}
The fixed prediction target is $g(u,v,z)=(u,z)$ with
\[
d_a[(u,z),(\tilde u,\tilde z)]
=\max\{d_{\Torus}(u,\tilde u),2a\,d_{\Torus}(z,\tilde z)\},
\qquad0<a<1/2.
\]
Take reference half-widths $r_u=r_v=r_z=1/2$.  Complete ignorance of the circle coordinate $z$ has minimax radius $a$ in this target metric.  Hence for $a\le\eps<1/2$ the $z$ coordinate need not be resolved.  The symmetric $u$-uncertainty through step $N$ has radius $\rho_u+N\rho_v$, so feasibility requires
\[
\rho_u+N\rho_v\le\eps.
\]
Maximizing the unresolved cell volume gives $\rho_u=\eps/2$, $\rho_v=\eps/(2N)$, and $\rho_z=1/2$.  Therefore
\[
B_N(\eps)=\log_2N+\cO_\eps(1).
\]

For $0<\eps<a$, $z$ must be resolved.  Before any wrap relevant to an admissible fine cell, the endpoint difference is
\[
\Delta z_N=m^N\Delta z_0+A_N\Delta u_0+C_N\Delta v_0,
\]
where
\[
A_N=\frac{m^N-1}{m-1},\qquad
C_N=\frac{m^N-Nm+N-1}{(m-1)^2}.
\]
For a rectangular initial cell the lifted $z$ displacements fill the interval $[-S_N,S_N]$, where
$S_N=A_N\rho_u+C_N\rho_v+m^N\rho_z$.  For $m\ge2$, the coefficients $A_n$, $C_n$, and $m^n$ are nondecreasing, so the endpoint $n=N$ gives the strongest $z$-channel constraint.  If $S_N\ge1/2$, reduction modulo one covers the entire $z$ circle and its minimax contribution is $a$, contradicting $\eps<a$.  Hence every feasible fine cell has $S_N<1/2$, so the lifted expression is exact and the $z$ endpoint constraint is

\[
A_N\rho_u+C_N\rho_v+m^N\rho_z\le\frac{\eps}{2a}=:L.
\]
Maximizing $\rho_u\rho_v\rho_z$ under this linear constraint gives equal contributions,
\[
\rho_u=\frac{L}{3A_N},\quad
\rho_v=\frac{L}{3C_N},\quad
\rho_z=\frac{L}{3m^N}.
\]
Because $A_N$ and $C_N$ grow exponentially, the accompanying $u$-channel constraint $\rho_u+N\rho_v\le\eps$ and the fixed reference-width caps are slack for all sufficiently large $N$.  The displayed allocation is therefore the exact optimum in that asymptotic range, and
\[
B_N(\eps)=\log_2(A_NC_Nm^N)+\cO(1)
=3N\log_2m+\cO(1).
\]
The slow and expanding channels are part of the same fixed dynamics and the same target; accuracy alone determines which one must be resolved.

\section{Forced Lorenz--84 model and two-scale diagnostic field}\label{app:lorenz84}
The numerical model is
\begin{align}
\dot X&=-Y^2-Z^2-aX+a(F_0+F_1\cos\theta),\\
\dot Y&=XY-bXZ-Y+G,\\
\dot Z&=bXY+XZ-Z,\\
\dot\theta&=c,\\
\dot c&=0,
\end{align}
with
\[
a=0.25,\quad b=4,\quad F_0=8,\quad F_1=0.5,\quad G=1,\quad c=0.25.
\]
The equation $\dot c=0$ includes the persistent external forcing parameter in the tangent calculation while leaving the atmospheric dynamics unchanged.  The target is
\[
H(\lambda,t)=A_P\cos[\lambda-\theta(t)]
+Y(t)\cos(k_e\lambda)+Z(t)\sin(k_e\lambda),
\qquad A_P=3,
\]
with $k_e=2$, the minimal harmonic distinct from the large-scale $k=1$ channel; any integer $k_e\ne1$ preserves the Fourier orthogonality used below.  With
\[
\|h\|_2=\left[\frac1\pi\int_0^{2\pi}|h(\lambda)|^2d\lambda\right]^{1/2},
\]
the three coefficient directions $\cos(\lambda-\theta)$, $\cos(k_e\lambda)$, and $\sin(k_e\lambda)$ are orthogonal.

For two finite trajectories,
\[
\|H_1-H_0\|_2=
\sqrt{4A_P^2\sin^2\!\left(\frac{\theta_1-\theta_0}{2}\right)
+(Y_1-Y_0)^2+(Z_1-Z_0)^2}.
\]
For an infinitesimal perturbation propagated by the tangent map $M(t)=D\bm q(t)/D\bm q(0)$, the coefficient norm generated by unit perturbation of initial coordinate $q_i$ is
\[
\gamma_i(t)=\sqrt{[A_PM_{\theta i}(t)]^2+M_{Yi}(t)^2+M_{Zi}(t)^2},
\]
and the horizon sensitivity is $\Gamma_i(T)=\sup_{0\le t\le T}\gamma_i(t)$.

\section{Coarse branch}\label{app:coarse}
Let $\mathcal K_{\rm atm}$ be the declared post-transient atmospheric admissible set and define
\[
\overline R_e:=\sup_{\mathcal K_{\rm atm}}\sqrt{Y^2+Z^2}.
\]
The coarse mechanism requires a uniform bound satisfying $\overline R_e<\eps_c<A_P$.  The numerical trajectory used here gives
\[
R_e^{\rm samp}=2.429159,
\]
with 99.5th percentile $2.209919$; $R_e^{\rm samp}$ is the empirical envelope of that sampled trajectory.  With $\eps_c=2.70$ and $A_P=3$, the sampled trajectory has a finite margin on both sides of the channel-activation threshold.

If the atmospheric eddy coordinates remain at their reference resolution, a zero $k_e=2$ estimate has worst-case eddy contribution bounded by $\overline R_e$.  For the external $k=1$ phase, let
\[
L_T=\rho_\theta+T\rho_c.
\]
For $L_T\le\pi/2$, the minimum enclosing radius in coefficient space of a phase arc with half-width $L_T$ is $A_P\sin L_T$.  Orthogonality therefore gives the sufficient condition
\[
\overline R_e^{\,2}+A_P^2\sin^2L_T\le\eps_c^2.
\]
Hence
\[
L_T\le \overline L_c:=\arcsin\frac{\sqrt{\eps_c^2-\overline R_e^{\,2}}}{A_P}.
\]
Using $R_e^{\rm samp}$ for the numerical curve gives $L_c=0.403760\ldots$.  Maximizing $\rho_\theta\rho_c$ under $\rho_\theta+T\rho_c\le L_c$ gives
\[
\rho_\theta=L_c/2,\qquad \rho_c=L_c/(2T).
\]
With reference widths $r_\theta=r_c=1$ the corresponding sufficient forcing-channel cost is
\[
B_c^{\rm suff}(T)=\log_2\frac{4T}{L_c^2}=\log_2T+\text{const.}
\]
Conversely, projection of any admissible prediction onto the orthogonal $k=1$ Fourier subspace removes the eddy mode.  Since $\eps_c<A_P$, the allowed phase-arc radius must remain bounded away from a full circle, which requires $\rho_\theta+T\rho_c=\cO(1)$ and therefore $\rho_c=\cO(T^{-1})$.  Thus, for any fixed admissible-set prior with a uniform eddy bound $\overline R_e<\eps_c$, the forcing channel has logarithmic precision cost from both sides.  The plotted Lorenz--84 coarse curve is the empirical realization obtained by substituting $R_e^{\rm samp}$ for $\overline R_e$.

\section{Fine linearized resource: lower and upper bounds}\label{app:fine}
For the fine branch all five coordinates are locally active.  Around each post-transient starting state, the uncertainty box is treated as a local refinement within the same declared admissible region.  Let $B^*_{\rm lin}(T;\bm q_0,\eps_f)$ be the exact minimum rectangular-box cost under the linearized horizon-wide target norm.  We use $d=5$, $\eps_f=0.35$, and reference widths $r_i=1$ in the declared model-coordinate units.

Varying only coordinate $i$ proves the necessary condition $\Gamma_i\rho_i\le\eps_f$.  Therefore
\[
B^*_{\rm lin}\ge
B_L:=\sum_i\left[\log_2\frac{r_i\Gamma_i}{\eps_f}\right]_+.
\]
The triangle inequality gives
\[
R^{\rm lin}_T(\bm\rho)\le\sum_i\Gamma_i\rho_i.
\]
The choice $\rho_i=\min[r_i,\eps_f/(d\Gamma_i)]$ is consequently feasible, so
\[
B^*_{\rm lin}\le
B_U:=\sum_i\left[\log_2\frac{d r_i\Gamma_i}{\eps_f}\right]_+.
\]
For each coordinate the two bracket contributions differ by at most $\log_2d$, hence
\[
0\le B_U-B_L\le d\log_2d=11.6096\ \text{bits}.
\]
The bound is independent of $T$, so $B_L$, $B_U$, and the exact local linearized optimum necessarily have the same asymptotic horizon class whenever either bound has a well-defined class.

\begin{figure}[htbp]
\centering
\includegraphics[width=0.72\linewidth]{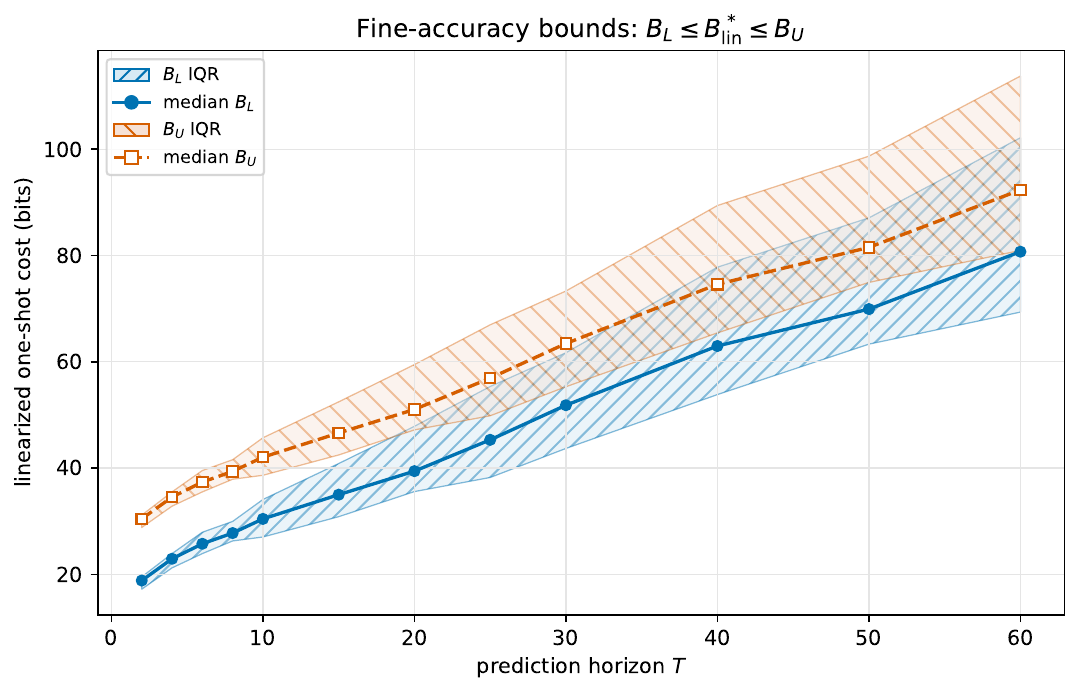}
\caption{Fine-accuracy local linearized resource bounds over 64 post-transient starting states.  The lower-bound median uses filled circles and a solid line; the upper-bound median uses open squares and a dashed line.  Oppositely hatched bands show the corresponding 25--75\% ranges, so the two bounds remain distinguishable in grayscale.  Their vertical separation is at most $d\log_2d$, independent of horizon.}
\label{fig:bounds}
\end{figure}

\begin{figure}[htbp]
\centering
\includegraphics[width=0.68\linewidth]{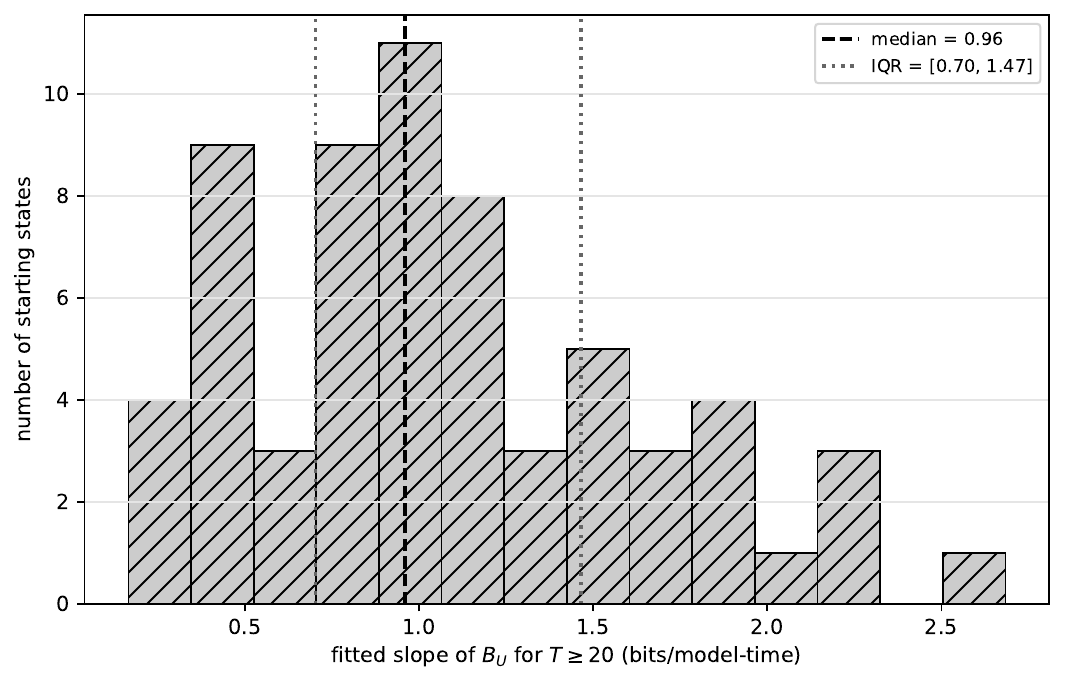}
\caption{Distribution of per-start fitted slopes of $B_U(T)$ over $T\ge20$ for the 64-state ensemble.  Hatched histogram bars, a dashed median line, and dotted interquartile markers provide grayscale-safe encodings.}
\label{fig:slopes}
\end{figure}

We integrated the state and full $5\times5$ tangent map with fourth-order Runge--Kutta at $\Delta t=0.01$.  Sixty-four post-transient starting states, sampled at separated times along one long reference trajectory, were used.  Figure~\ref{fig:bounds} shows the lower/upper brackets and their initial-condition variability.  Over $T\ge20$, the median $B_U$ curve has fitted slope $1.0125$ bits per model-time unit.  Fitting each starting state separately gives a slope median $0.9598$ and interquartile range $[0.7034,1.4655]$.  The broad per-start distribution records genuine finite-time dynamical variability across the sampled starting states.  Figure~\ref{fig:slopes} displays that variability explicitly: most fitted slopes cluster near the ensemble-median behavior, while the broad tail shows that the finite-window growth rate depends appreciably on the starting state.

\section{Direct nonlinear finite perturbations}\label{app:direct}
To check that the tangent result is not solely a linearization artifact over the resolved window, we directly evolved finite trajectory pairs.  Sixteen post-transient base states, 20 perturbation directions (10 signed coordinate directions and 10 fixed-seed mixed directions), and 48 logarithmically spaced perturbation magnitudes were used.  At each horizon the exact two-scale field distance above was maximized over sampled bases, directions, and all prior times and then inverted at $\eps_f=0.35$.

\begin{figure}[htbp]
\centering
\includegraphics[width=0.70\linewidth]{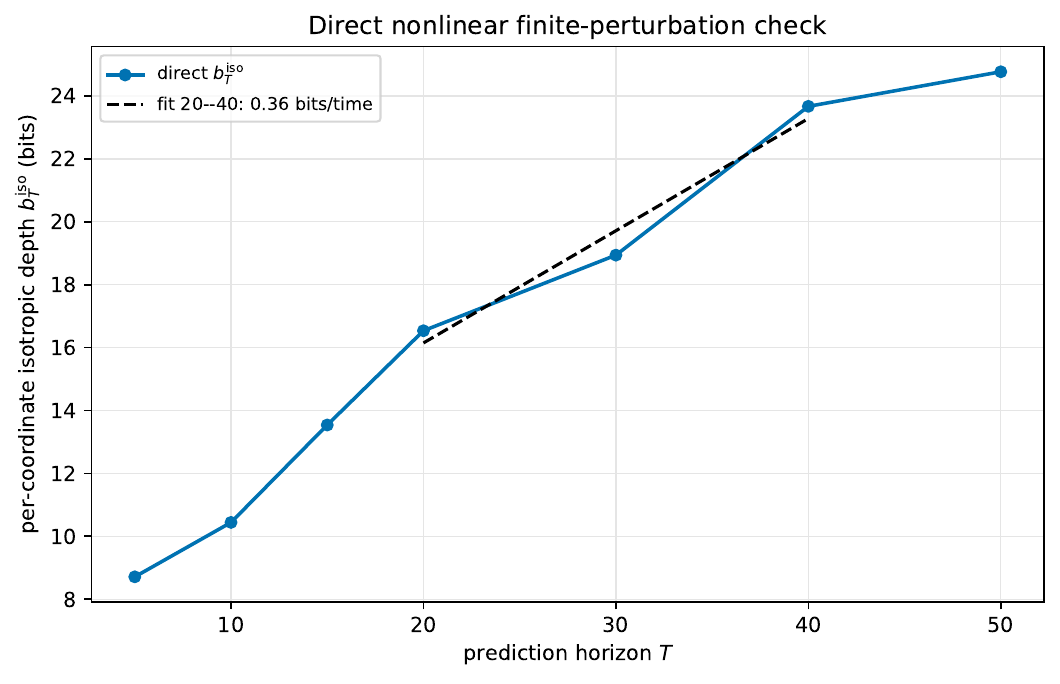}
\caption{Direct nonlinear finite-perturbation check.  Filled circles joined by a solid line show $b_T^{\rm iso}$, while the black dashed segment is the $20\le T\le40$ fit.  The ordinate is the per-coordinate isotropic depth, not the total anisotropic resource $B_T$.}
\label{fig:direct}
\end{figure}

This isotropic calculation measures a complementary quantity to the anisotropically optimized total resource $B_T$: the per-coordinate isotropic precision depth
\[
b_T^{\rm iso}:=-\log_2\delta_T^*,
\]
where $\delta_T^*$ is the largest sampled normalized equal-width perturbation satisfying the tolerance.  The results are
\[
\begin{array}{c|rrrrrrr}
T&5&10&15&20&30&40&50\\\hline
b_T^{\rm iso}&8.71&10.44&13.54&16.54&18.94&23.67&24.77.
\end{array}
\]
The fit over $20\le T\le40$ gives $0.357$ bits per coordinate per model-time unit.  The longest horizons enter the resolution limit of the sampled amplitude/direction grid.  The informative window is therefore the pre-saturation range, where the direct nonlinear calculation independently shows an increasing, approximately linear precision depth.  This numerical saturation limits the verification window rather than indicating saturation of the underlying predictive resource.  Figure~\ref{fig:direct} shows both the pre-saturation growth and the subsequent numerical flattening, thereby separating the nonlinear evidence for increasing precision demand from the sampling limit of this check.

\section{Parameter robustness}\label{app:scan}
\begin{figure}[htbp]
\centering
\includegraphics[width=0.72\linewidth]{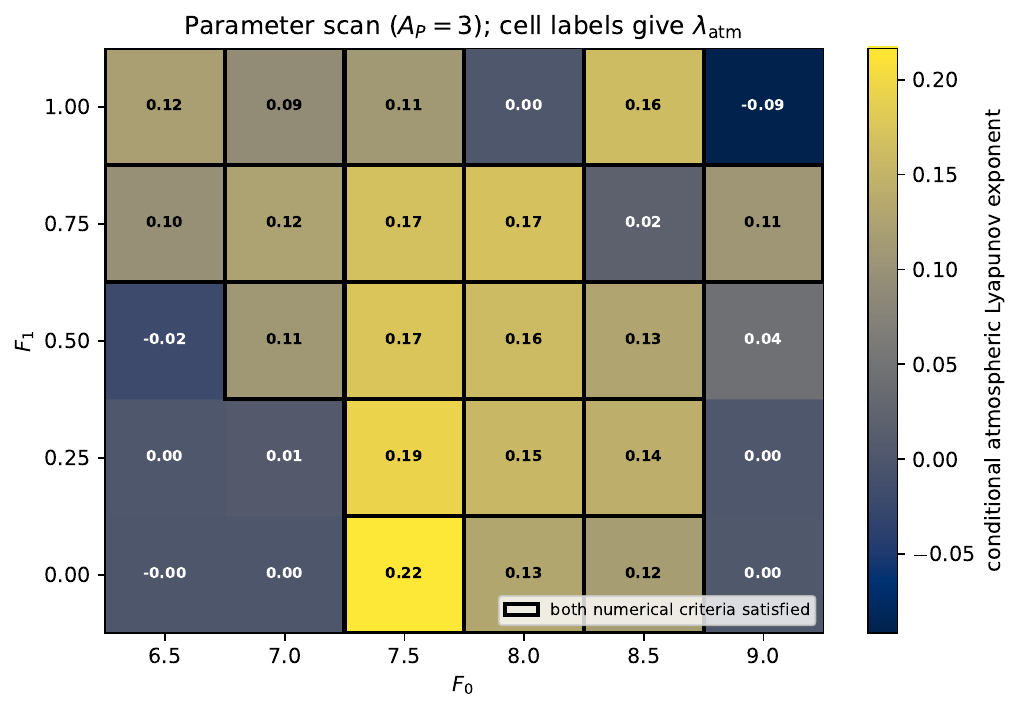}
\caption{Parameter scan of the forced Lorenz--84 dynamics.  Each cell is labeled by the numerical conditional atmospheric Lyapunov exponent, and a thick black cell border marks points satisfying both the positive-instability and coarse-amplitude-margin criteria.  The monotonic-luminance background is auxiliary; the numerical labels and borders preserve the information in grayscale.}
\label{fig:scan}
\end{figure}

We varied the forcing parameters to test whether positive atmospheric instability and a coarse-amplitude margin persist beyond the reference point.  We scanned
\[
F_0\in\{6.5,7.0,7.5,8.0,8.5,9.0\},\qquad
F_1\in\{0,0.25,0.5,0.75,1.0\},
\]
at $c=0.25$.  All 30 sampled points had $R_e^{\rm samp}<2.70$, leaving a nonempty coarse interval for the chosen $A_P=3$ and $\eps_c=2.70$.  Nineteen of 30 also had a numerically positive conditional atmospheric Lyapunov exponent exceeding $0.05$ together with a coarse-amplitude margin above $0.20$; Fig.~\ref{fig:scan} marks the points satisfying both numerical criteria.  The scan shows that the channel-switching conditions are not confined to a single parameter choice.

At $F_0=8$, $F_1=0.5$, scanning $c=0.15,0.20,0.25,0.30,0.35,0.40$ preserved positive conditional atmospheric exponents of roughly $0.14$--$0.18$ and a finite eddy-amplitude margin in the numerical runs.

\section{Autonomous closure}\label{app:autonomous}
\begin{figure}[htbp]
\centering
\includegraphics[width=0.72\linewidth]{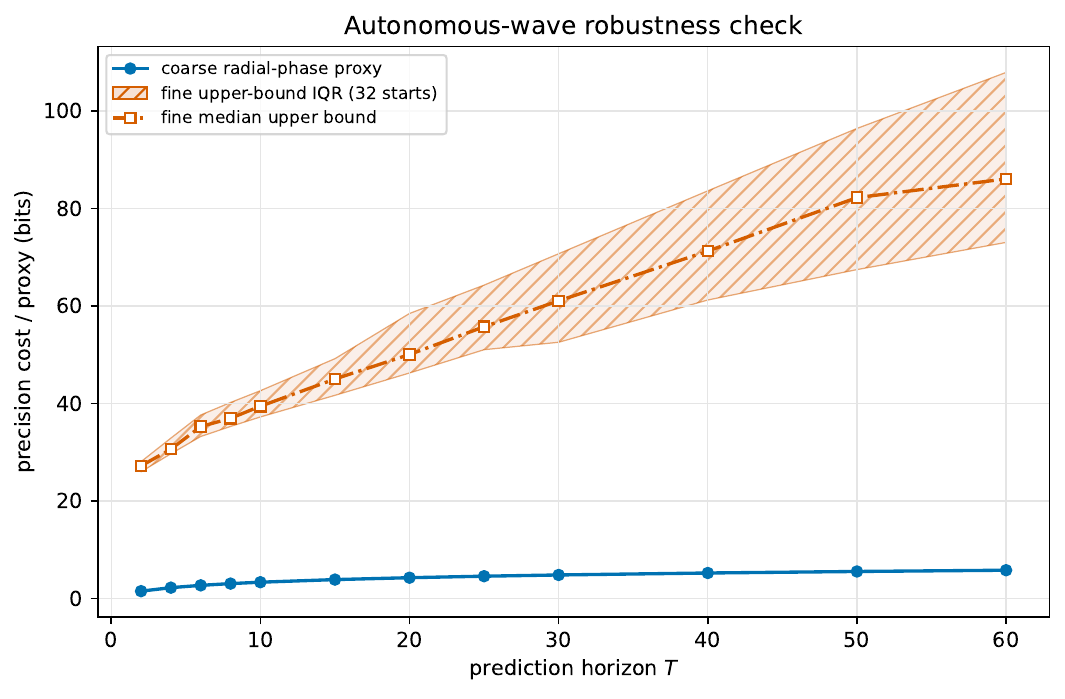}
\caption{Autonomous-wave robustness calculation using the same orthogonal two-scale diagnostic structure.  Filled circles with a solid line show the coarse radial-phase asymptotic proxy; open squares with a dash-dotted line show the fine median upper bound, and the hatched band gives its 25--75\% initial-condition range over 32 starts.  These encodings remain distinct in grayscale.}
\label{fig:autonomous}
\end{figure}

The external forcing description is useful because its phase rate can be treated as an estimated or controlled external quantity.  To test whether the mechanism depends on a frozen uncertain parameter, we also close the forcing dynamics autonomously.  Let $W=P+iQ$ obey
\[
\dot W=i\Omega(|W|^2)W,
\qquad
\Omega(R^2)=\omega_0+\kappa(R^2-R_0^2),
\]
with $R_0=3$, $\omega_0=0.25$, and $\kappa=0.05$.  It drives Lorenz--84 through $F=F_0+F_1P/R_0$.  The orthogonal diagnostic field is now
\[
H(\lambda,t)=P(t)\cos\lambda+Q(t)\sin\lambda+Y(t)\cos(2\lambda)+Z(t)\sin(2\lambda).
\]
At nominal radius $R_0$,
\[
\left|\frac{\partial W(t)}{\partial R}\right|
=\sqrt{1+4\kappa^2R_0^4t^2}
\sim2|\kappa|R_0^2t,
\]
so radial state uncertainty alone generates a $T^{-1}$ coarse resolution requirement and logarithmic cost.  In 32 post-transient starts, the fine upper-bound median for the orthogonal diagnostic field had fitted slope $0.94$ bits per model-time unit for $T\ge20$.  The sampled eddy maximum was $2.4364$ and the conserved wave-radius drift remained below $5\times10^{-12}$ in the calculation.  Figure~\ref{fig:autonomous} shows that the same coarse--fine separation persists after replacing the frozen forcing parameter by autonomous wave dynamics, supporting the interpretation as a channel-activation mechanism rather than an artifact of external-parameter bookkeeping.

Direct finite-perturbation calculations for the autonomous closure likewise show increasing per-coordinate isotropic precision depth over the resolved pre-saturation window, providing a second finite-amplitude check of the mechanism.

\section{Operational interpretation}\label{app:scope}
The exact and numerical parts of the paper play different roles.  The toral construction proves the asymptotic $\Th(\log T)\to\Th(T)$ accuracy-selected class switch for one fixed dynamics and target.  The Lorenz--84 calculation shows how the same channel-activation mechanism appears in a coupled continuous system over finite resolved horizons: the coarse forcing-only branch follows from a uniform eddy-amplitude prior, while the fine branch locally refines all five coordinates and exposes exponential atmospheric sensitivity.  The ensemble spread in Figs.~\ref{fig:scaling}--\ref{fig:slopes} is part of that finite-time dynamics, and the direct finite-perturbation calculation provides an independent nonlinear check over its pre-saturation window.

The target field is a single scalar two-scale diagnostic with orthogonal Fourier channels.  Orthogonality separates the large-scale forcing error from the intrinsic eddy coefficients in the prediction norm; the mechanism itself is the activation of target channels with different horizon-growth laws.  Likewise, the coordinate system is part of the sensing architecture: fixed rescalings change only horizon-independent offsets, while a different coordinate mixing represents a different allocation problem.

This one-shot resource complements finite-size Lyapunov analysis, targeted observations, estimation entropy, restoration entropy, and data-rate theorems.  Those constructions describe sensitivity, observation targeting, orbit reconstruction, or continuing information rates.  Here the output is instead a joint prediction-design statement: how much initial precision is required, how it is allocated among declared physical channels, and how that minimum grows with horizon.  In a sensing-design setting, this allocation can identify which state or forcing channels require additional resolution for a specified forecast horizon and tolerance, thereby informing sensor bit-depth and observation-budget design.  The exact transition demonstrates that the accuracy requirement itself can select the governing resource law.

\section*{Acknowledgments}
This work was supported by the research grant SP015 from SOBIN Institute LLC.

\end{document}